# Comparison of different Propagation Steps for the Lattice Boltzmann Method

Markus Wittmann[a], Thomas Zeiser[a], Georg Hager[a], Gerhard Wellein[b]

[a]*Erlangen Regional Computing Center (RRZE), University of Erlangen-Nuremberg, Germany*
[b]*Department of Computer Science, University of Erlangen-Nuremberg, Germany*

**Abstract**

Several possibilities exist to implement the propagation step of the lattice Boltzmann method. This paper describes common implementations which are compared according to the number of memory transfer operations they require per lattice node update. A memory bandwidth based performance model is then used to obtain an estimation of the maximum reachable performance on different machines. A subset of the discussed implementations of the propagation step were benchmarked on different Intel and AMD-based compute nodes using the framework of an existing flow solver which is specially adapted to simulate flow in porous media. Finally the estimated performance is compared to the measured one. As expected, the number of memory transfers has a significant impact on performance. Advanced approaches for the propagation step like "AA pattern" or "Esoteric Twist" require more implementation effort but often sustain significantly better performance than non-naïve straightforward implementations.

## 1. Introduction

The lattice Boltzmann method (LBM) [1] is used in computational fluid dynamics to simulate incompressible flows. Physically, the discrete time step in a LBM simulation consists of collision and propagation. In the collision step the particle collisions are modeled. The *propagation step* streams the new information to neighboring nodes. This paper gives an overview on several techniques that can be used to accomplish propagation and discusses how to optimize propagation with respect to performance.

Different stencil types can be applied for the lattice Boltzmann method. In general they are named D$d$Q$q$, where $d$ describes the dimension (two or three dimensional) and $q$

---

*Email addresses:* markus.wittmann@rrze.uni-erlangen.de (Markus Wittmann),
thomas.zeiser@rrze.uni-erlangen.de (Thomas Zeiser), georg.hager@rrze.uni-erlangen.de (Georg Hager), gerhard.wellein@rrze.uni-erlangen.de (Gerhard Wellein) denominates the number of particle distribution functions (PDFs). Typical examples are D2Q9 or D3Q19.

The propagation step is usually independent of the underlying lattice type as data always has to be streamed to neighboring nodes. However, depending on how nodes are allocated in memory, neighboring nodes can be determined either by simple index arithmetic ("direct addressing of a full grid") or by indirect addressing [2, 3] ("sparse, pseudo-unstructured representation of the underlying Cartesian grid"; an index array "IDX" holds the full adjacency information of all nodes, i.e. the $q-1$ neighbors per node). In this paper we discuss both, direct and indirect addressing. The PDFs of both lattice types can be represented by a propagation-optimized *structure of arrays* (SoA) data layout where each direction of the discrete velocities is stored in its own array, or by a collision-optimized *array of structures* (AoS) data layout. With AoS only one array exists and the PDFs are stored node-wise. Typically, SoA results in better cache utilization without reverting to loop blocking or similar techniques [10]. The benchmark measurements discussed in Section 10 use a framework which implements indirect addressing and the SoA data layout.

Memory bandwidth is a major bottleneck of all modern general purpose computers. Hierarchies of caches have been introduced since many years to mitigate the limited memory transfer rate. Data typically has to be moved from and to main memory in chunks of *cache lines* (128 bytes). Data in main memory cannot be modified directly, but the corresponding cache line has to be loaded from main memory to cache first. Finally, the cache line will be evicted back to main memory [13]. This three-step procedure occurs automatically and fully transparently. The process of loading data to cache to allow modification is called *write allocate* or *read for ownership (RFO)* and ensures consistency of data in all memory domains and hierarchies of multi-core/multi-socket compute nodes. *Non-temporal stores* are a special case as they bypass the cache hierarchy and directly write to main memory.

This paper is organized as follows: First the basic *two-step* algorithm is introduced in Section 2 where collision and propagation are handled separately in two steps. The *one-step* algorithm in Section 3 fuses the collision and propagation into one step. Both previous algorithms usually use two lattices to easily avoid data dependencies. As memory is valuable, algorithms were developed which elude the data dependencies and thus can go with one lattice. The *compressed grid* algorithm of POHL et al. [4] in Section 4 shifts values back and forth in one lattice between time steps. The *swap algorithm* of MATTILA et al. [5] and LATT [6] in Section 5 uses a strict processing order and swapping of PDFs to circumvent data dependencies. BAILEY et al.'s *AA-pattern* [7] discussed in Section 6 requires two different lattice Boltzmann kernels for updating the single lattice. Section 7 evaluates the *esoteric twist* algorithm of GEIER et al. [8, 9], which uses pointer swapping after each time step. Our test bed for the benchmark measurements is introduced in Section 8, and a performance model including estimates for the different propagation models is presented in Section 9. A set of these algorithms were implemented and achieved performance is compared to predictions in Section 10. The summary in Section 11 concludes the paper.

*Related Work.* POHL et al. [4] compared an implementation of the one-step algorithm with the compressed grid technique. In [10], WELLEIN et al. analyzed different data layouts in combination with the one-step algorithm on a wide range of different architectures. MAT-TILA et al. compared in [5] implementations of the two-step, one-step, compressed grid, swap, and Langrangian algorithms with direct, semi-direct, and indirect addressing.

|  | direct addressing | | | indirect addressing | |
|---|---|---|---|---|---|
| # | type | comment | # | type | comment |
|  |  | *collision* |  |  | *collision* |
| $q$ | PDF | loads $A$ | $q$ | PDF | loads $A$ |
| $q$ | PDF | write allocate $B$ | $q-1$ | IDX | stores into $B$ |
| $q$ | PDF | stores $B$ | $q$ | PDF | write allocate $B$ |
|  |  |  | $q$ | PDF | stores $B$ |
|  |  | *propagation* |  |  | *propagation* |
| $q$ | PDF | loads $B$ | $q$ | PDF | loads $B$ |
| $q$ | PDF | write allocate $A$ | $q-1$ | IDX | stores into $A$ |
| $q$ | PDF | stores $A$ | $q$ | PDF | write allocate $A$ |
|  |  |  | $q$ | PDF | stores $A$ |
| $\Sigma = 6q * \text{PDFs}$ | | | $\Sigma = 6q * \text{PDFs} + 2(q-1) * \text{IDXs}$ | | |

Table 1: Estimating the data requirements for the **two-step** algorithm for one complete lattice node update.

Using temporal blocking (e.g. [4]) or wavefront techniques (e.g. [11]) are another approach to reduce the required memory bandwidth. However, all these latter approaches which exploit the increase of temporal locality by fusing several time steps are beyond the scope of this paper.

## 2. Two-Step Two-Grid Algorithm

The two-step algorithm (TS) is a naïve approach taking collision and propagation literally as two separate steps. In the worst case two lattices $A$ and $B$ are used where A holds the PDFs of the nodes (often denoted as $f_i$) and B the intermediate post-collision pre-propagation values (often denoted $f_i^*$). For each node in lattice $A$ the collision is performed and the newly computed PDFs are stored at the same spatial location but in array $B$. The following propagation step (i.e. a different loop over all nodes) streams the post-collision values from array $B$ of a node to its neighbor back into lattice $A$ to have the PDFs of the next time step.

This algorithm is very memory bandwidth intensive as data has to be transferred several times in each time step over the memory bus. An overview of the memory transfers for one lattice node update is given in Tab. 1.

## 3. One-Step Two-Grid Algorithm

In the one-step algorithm (OS) collision and propagation are fused. For a node in lattice $A$ the collision is performed (Fig 1(a)). The post-collision values of this node are immediately streamed to their neighbors and stored in lattice $B$ (Fig 1(b)). After all nodes have been updated, lattice pointers to $A$ and $B$ are swapped and the iteration starts over. Table 2 summarizes the memory accesses for a single node update.

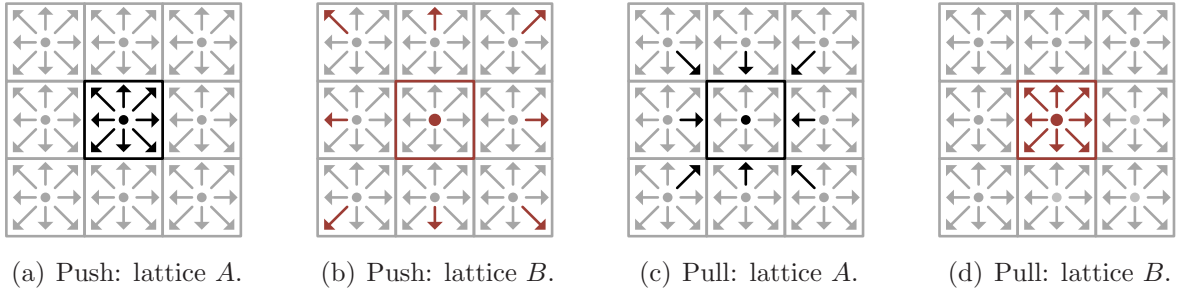

| (a) Push: lattice $A$. | (b) Push: lattice $B$. | (c) Pull: lattice $A$. | (d) Pull: lattice $B$. |

Figure 1: The **one-step** algorithm for **push** and **pull** scheme. Particle distribution functions (PDFs) are read from lattice $A$ (a) and after collision they are written to lattice $B$ (b). This scheme is also known as push whereas for pull the propagation (c) takes place before the collision (d).

| direct addressing | | | indirect addressing | | |
|---|---|---|---|---|---|
| # | type | comment | # | type | comment |
| $q$ | PDF | loads $A$ | $q$ | PDF | loads $A$ |
| $q$ | PDF | write allocate $B$ | $q-1$ | IDX | stores into $B$ |
| $q$ | PDF | stores $B$ | $q$ | PDF | write allocate $B$ |
| | | | $q$ | PDF | stores $B$ |
| $\Sigma = 3q * \text{PDFs}$ | | | $\Sigma = 3q * \text{PDFs} + (q-1) * \text{IDXs}$ | | |

Table 2: Estimating the data requirements for the **one-step** algorithm. The table shows only the store and load order for **push** scheme, which is different to the *pull* scheme. Despite the actual number of memory accesses and accesses types is the same.

Performing collision before propagation is referred to as the *push* scheme. The post-collision values are scattered to their neighbors. Swapping the two steps, i.e. arranging propagation before collision, is called *pull* scheme and data has to be gathered from the neighbors. Both schemes can be used together with the one-step algorithm resulting in the same number of memory accesses.

Performing collision and propagation together requires only loading from $A$ and storing to $B$ once, which reduces the required memory bandwidth for a node update by 50 % compared to the worst-case two-step two-grid algorithm.

*Non-temporal stores.* Push and pull scheme only write to array $B$. With non-temporal stores, which bypass the cache hierarchy, the write allocate can be avoided, reducing the overall memory traffic by 1/3. A particularly efficient implementation (OS-NT) is *OS with pull and SoA* as SIMD-vectorized instructions and non-temporal stores can be used to write to consecutive memory. All stores must now be performed on addresses which are a multiple of 16 (for SSE). Since at least two elements are written at once, the SoA data layout is required. For direct addressing each line in the lattice must be correctly aligned and thus may need padding. In case of indirect addressing only each array (i.e. the $q$ arrays of the $q$ direct velocity directions) must be correctly aligned.

Modern x86 architectures can only sustain a small number of concurrent non-temporal store streams without incurring a performance penalty. To reduce the number of concurrent write streams, collision/propagation has to be spatially blocked using chunks of a certain length $l$, e.g., 150 nodes. The original PDFs and some calculated values (e.g., the macroscopic values of one block) are first transferred to a temporary array to hold the data of the $l$ nodes in the innermost cache. In $q$ (OS-NT, Pull, 1S) or $q/2$ (OS-NT, Pull, 2S) subsequent loops, collision and propagation takes place by fetching the required input data from the temporary array (which still is in cache), doing the calculations, and writing the post-collision values back to lattice $B$ using non-temporal stores. Using pairs of opposing directions is particularly useful for TRT as they share common sub-expressions; however, compilers may generate write streams with only one non-temporal store. Thus, it may be required to use $q$ loops with length $l$ each to have exactly one concurrent write stream only.

Bandwidth requirements for one node are reduced for the full array type and indirect addressing by $q * sizeof(PDF)$ bytes, which results in $2 * q * sizeof(PDF)$ bytes/node update for the full array and $2 * q * sizeof(PDF) + (q-1) * siezof(IDX)$ bytes/node for indirect addressing. Accesses to the temporary array are neglected in our simple performance model as it is so small that it can be kept in L1 cache and it is assumed that accessing L1 cache is orders of magnitude faster than accessing memory. The assumption of infinitely fast L1 access may not be correct for all architectures (in particular AMD Opterons with their exclusive caches).

## 4. Compressed Grid

The compressed grid technique (CG) of POHL et al. [4], also known as the *shift algorithm* [7] puts the two lattices previously needed on top of each other and shifted by one cell in each direction. Thus, only one lattice has to be stored, plus one layer of cells in each dimension. The data dependency between the nodes is broken by writing updated PDFs to the lattice node shifted by the vector **s**. For even time steps the iteration takes place from top right to bottom left as in Fig. 2(a) and $\mathbf{s} = \mathbf{s}_e = (1,1)^T$. In odd time steps, iteration is performed from the bottom left to top right as in Fig. 2(b) with a shift vector of $\mathbf{s} = \mathbf{s}_o = (-1,-1)^T$. Although the description is given for the 2D case this can be easily extended to 3D. Details about the memory transfers involved are provided in Tab. 3.

Only one kernel is needed, with the update order of the lattice nodes alternating between successive time steps. Depending on the compiler, it might be beneficial to have a separate loop for each direction so that the SIMD vectorizer has all the needed information. Compressed grid can be used with the push and the pull scheme.

## 5. Swap Algorithm

The swap algorithm (SWAP) of MATILLA et al. [5] and LATT [6] falls also into the class of propagation methods which require only one lattice. Data dependencies are circumvented by a strict processing order of the lattice nodes and by continuously swapping half of a node's PDFs with its neighbors'. Iteration over the lattice nodes takes place in lexicographic order. Which PDFs to swap depends on whether the push or pull scheme is used.

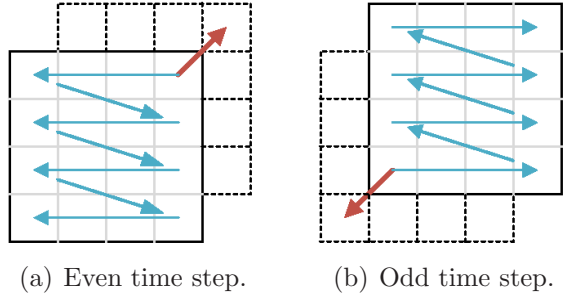

(a) Even time step.　　(b) Odd time step.

Figure 2: Schematic sketch of the **compressed grid** technique requiring only on lattice. Iteration order is alternated between two time steps. Together with a lattice increased by one cell in each dimension the data dependencies are broken.

| direct addressing | | | indirect addressing | | |
|---|---|---|---|---|---|
| # | type | comment | # | type | comment |
| $q$ | PDF | loads | $q$ | PDF | loads |
| $q-1$ | PDF | write allocate | $q-1$ | IDX | loads |
| $q$ | PDF | stores | $q-1$ | PDF | write allocate |
| | | | $q$ | PDF | stores |
| $\Sigma = (3q-1) * \text{PDFs}$ | | | $\Sigma = (3q-1) * \text{PDFs} + (q-1) * \text{IDXs}$ | | |

Table 3: Estimating the data requirements for one node update with **compressed grid**.

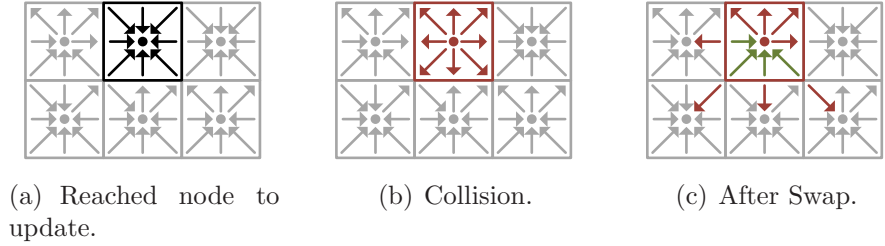

(a) Reached node to update.   (b) Collision.   (c) After Swap.

Figure 3: The **swap** algorithm for the **push** scheme. Nodes which have not been updated in the current iteration have their PDFs stored at opposing locations (black arrows in (a)). Post-collision PDFs are written back to their natural location (red arrows in (b)). All PDFs pointing to neighbor nodes that have already been visited in this iteration are swapped with those neighbor PDFs pointing to them (green arrows in (c)).

In case of the push scheme, PDFs of nodes that have not yet been visited in the current iteration are stored at the "opposing" location (black arrows in Fig. 3(a)), i.e. pointing to the center PDF of a node. When a node is reached (Fig. 3(a)) its PDFs are read and collided. The post-collision PDFs are written back to their "natural" location (red arrows in Fig. 3(b)). All PDFs that now point to neighbor nodes that have already been visited are swapped with the neighbor PDFs pointing to them (Fig. 3(c)).

In the swap algorithm the propagation step is implicitly performed by swapping PDFs. Special care has to be taken if data is processed after the end of the time step as the "PDFs of a node" are not all local.

After the whole lattice has been updated, PDFs of nodes at the boundaries which have not been swapped have to be swapped. At the beginning of the iteration, PDFs which are located at the boundary of the lattice must also be swapped or manually initialized. In general, blocking can be performed, but it generates additional overhead as the PDF values at block boundary must be swapped manually.

Using pull instead of the push scheme is also possible. Here, the PDFs of the current node (black arrows in Fig. 4(a)) have to be swapped first. The PDFs pointing to neighbor nodes that have not already been visited are swapped with the neighbor PDFs pointing to them (green arrows) as in Fig. 4(b). Now all PDFs needed for collision are located at the local node. Note that the PDFs are located at opposing locations. These PDFs are read and collided. All post-collision PDFs are written back and stored in their natural position in the local node (red arrows in Fig. 4(c)).

The required transfers and the different load/store order for both schemes can be found in Tab. 4.

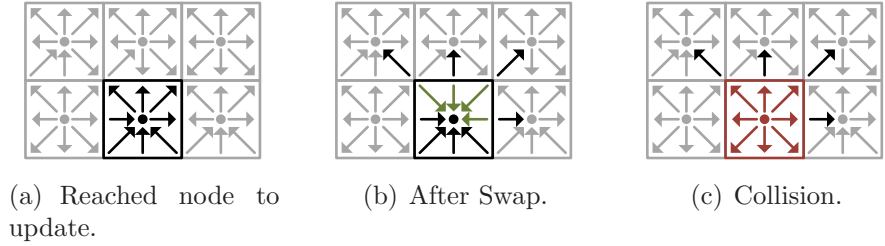

(a) Reached node to update.

(b) After Swap.

(c) Collision.

Figure 4: The **swap** algorithm for the **pull** scheme. For a node to update (black arrows in (a)) first the swap is performed. PDFs of this node, pointing to neighbor nodes which have not already been visited are swapped with those neighbor PDFs pointing to them (green arrows in (b)). After the PDFs of the local node have been read and collided the post-collision PDFs (red arrows) are written back to their natural position (c).

| direct addressing | | | indirect addressing | | |
|---|---|---|---|---|---|
| # | type | comment | # | type | comment |
| *push scheme* | | | | | |
| | | *collision* | | | *collision* |
| $q$ | PDF | loads | $q$ | PDF | loads |
| $(q+1)/2$ | PDF | stores local | $(q+1)/2$ | PDF | stores local |
| $(q-1)/2$ | PDF | load remote for swap | $(q-1)/2$ | IDX | for swap with remote |
| $q-1$ | PDF | store for swap | $(q-1)/2$ | PDF | load remote for swap |
| | | | $q-1$ | PDF | stores for swap |
| *pull scheme* | | | | | |
| $q$ | PDF | loads | $q$ | PDF | loads |
| $(q-1)/2$ | PDF | load remote for swap | $(q-1)/2$ | IDX | for swap with remote |
| $(q-1)/2$ | PDF | store remote for swap | $(q-1)/2$ | PDF | load remote for swap |
| $q$ | PDF | stores local | $(q-1)/2$ | PDF | store remote for swap |
| | | | $q$ | PDF | stores local |
| $\Sigma = (3q-1) * \text{PDFs}$ | | | $\Sigma = (3q-1) * \text{PDFs} + (q-1)/2 * \text{IDXs}$ | | |

Table 4: Estimation of data requirements for one node update with the **swap** algorithm using the **push** or **pull** scheme.

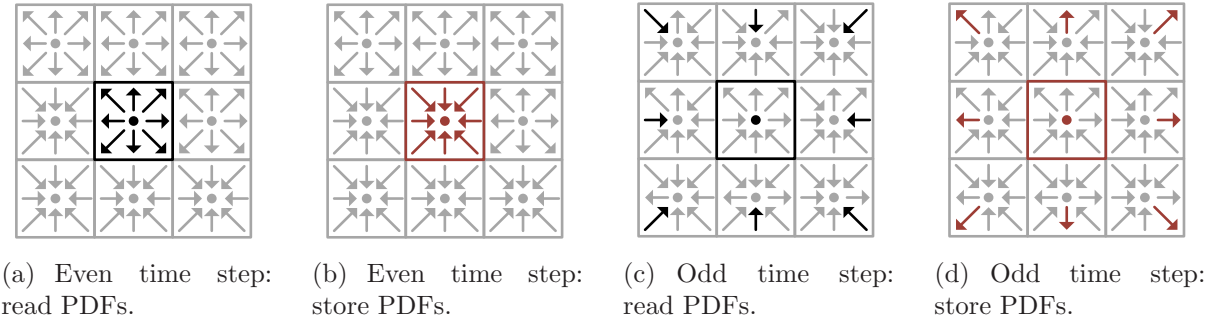

(a) Even time step: read PDFs.
(b) Even time step: store PDFs.
(c) Odd time step: read PDFs.
(d) Odd time step: store PDFs.

Figure 5: Schematic sketch of the **AA pattern**. At even time steps only node-local PDFs are accessed. They are read (black arrows in (a)), collided, and post-collision values (red arrows) are written back to their opposing position (b). At odd time steps only PDFs of neighbor nodes are accessed (black arrows in (c)), except the rest particle. After they have been read and collided, they are written back so they point away of the local node's center (red arrows in (d)).

## 6. AA Pattern

The AA pattern (AAP) of BAILEY et al. [7] requires only one lattice similar to the swap algorithm, but the processing order of the lattice nodes is not restricted in any way. The algorithm is inherently parallel, which means that nodes can be updated independently of each other. This makes it attractive for general purpose GPUs (GPGPUs), where memory capacity is low and massive parallelism is required to achieve reasonable performance. However, the algorithm is more complex and consists of different access patterns in even and odd time steps.

At even time steps only collisions are performed. PDFs of lattice nodes which have not been updated are in their natural position (black arrows) as shown in Fig. 5(a). For collision, PDFs of one node are read, collided, and post-collision PDFs (red arrows) are written back to the same node but at their opposing location Fig. 5(b), i.e. their orientation is locally swapped.

Odd time steps act like propagation – collision – propagation. PDFs from neighbors pointing to the current node (shown as black arrows in Fig. 5(c)) are read (first propagation). After they haven been collided the post-collision PDFs (red arrows) are stored at the previously read locations, but now again in the natural orientation (second propagation with rotation, Fig. 5(d)). An overview of the data requirements is given in Tab. 5.

These two different alternating steps require two different versions of all routines dealing with the PDFs in the lattice. At least the collide-propagate kernel and the in-/outlet routines are affected.

A really nice feature from the hardware point of view is that during the same update step writes are performed only to locations that were previously read. This eliminates the write allocate in the cache hierarchy, because the needed cache line to store to is already present. So each time step whether it is even or odd only requires $q$ loads and $q$ stores only.

| direct addressing | | | indirect addressing | | |
|---|---|---|---|---|---|
| # | type | comment | # | type | comment |
| | | *even time-steps* | | | *even time-steps* |
| $q$ | PDF | loads | $q$ | PDF | loads local |
| $q$ | PDF | stores | $q$ | PDF | stores local |
| | | *odd time-steps* | | | *odd time-steps* |
| $q$ | PDF | loads | $q$ | PDF | loads remote |
| $q$ | PDF | stores | $q$ | PDF | store remote |
| | | | $(q-1)$ | IDX | access neighbors |
| $\Sigma = 2q * \text{PDFs}$ | | | $\Sigma = 2q * \text{PDFs} + (q-1)/2 * \text{IDXs}$ | | |

Table 5: Estimation of data requirements for one node update with the **AA pattern**. Direct addressing requires always $q$ loads and $q$ stores no matter whether the time step is even or odd. With indirect addressing for odd time steps where PDFs of neighbors are accessed additional $q-1$ look-ups in the adjacency list (IDX) are required. These additional loads are equally distributed over both types of time steps for the final sum of bytes required per node update.

## 7. Esoteric Twist

The esoteric twist algorithm (ET) of GEIER et al. [8, 9] shares some characteristics with BAILEY's AA pattern, like the single lattice requirement and that it is intrinsically parallel. However this algorithm requires only one incarnation of the LB kernel, in-/outlet, etc. functions, which is beneficial when development time and maintainability of the code matters. The fact that only one lattice is needed and nodes can be updated independently of each other makes it again a good candidate for the usage on GPGPUS.

Independently whether direct or indirect addressing is used, a structure of array (SoA) data layout for accessing the PDFs is required, where the control structure contains pointer to the arrays of the different PDF directions.

With esoteric twist, the PDFs belonging to one node are only the half centered around the node. The rest are the PDFs pointing to the node's local PDFs. These are the black arrows in Fig. 6(a). The PDFs are read, collided and written back to the opposite direction as Fig. 6(b) shows. After each node of the lattice has been updated, the pointers of opposing discrete velocity directions in the control structure are exchanged. These pointer swaps not only recreates the shape of the lattice to the one before the update. In fact, it performs the propagation as Fig. 6(c) depicts. Details on the transferred data are found in Tab. 6.

Only writes to locations from where previously was read from occur in the algorithm. This eliminates additional loads for the write allocate of the cache hierarchy. The swap of pointers at the end of each iteration is negligible.

## 8. Test Bed

As test bed for our quantitative performance predictions and the performance measurements we used four different machines. Details of the systems can be found in Tab. 7.

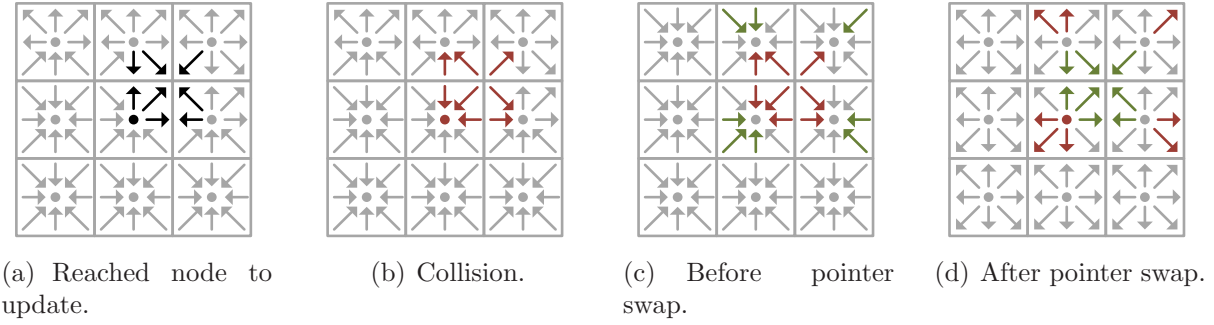

(a) Reached node to update.  (b) Collision.  (c) Before pointer swap.  (d) After pointer swap.

Figure 6: Schematic sketch of the **esoteric twist**. For updating a node its local and remote PDFs (indicated by the black arrows) are read (a), collided, and written back to opposing directions (red arrows in (b)). After each node has been updated in that way (c), the pointers of opposing PDF directions in the control structure are swapped, which implicitly performs the propagation (d).

| direct addressing | | | indirect addressing | | |
|---|---|---|---|---|---|
| # | type | comment | # | type | comment |
| $q$ | PDF | loads | $(q+1)/2$ | IDX | access neighbor links |
| $q$ | PDF | stores | $q$ | PDF | loads |
| | | | $q$ | PDF | stores |
| $\Sigma = 2q * \text{PDFs}$ | | | $\Sigma = 2q * \text{PDFs} + (q+1)/2 * \text{IDXs}$ | | |

Table 6: Data requirements for one node update with the **esoteric twist**.

|                                 | **Harpertown** | **Westmere** | **SandyBridge** | **MagnyCours** |
|---------------------------------|:---:|:---:|:---:|:---:|
| Name                            | Intel Xeon X5482 | Intel Xeon X5650 | Intel Xeon E3-1280 | AMD Opteron 6134 |
| Freq. [GHz]                     | 3.20 | 2.67 | 3.50 | 2.30 |
| max [GHz]                       | – | 3.06 | 3.90 | – |
| Cores                           | $2 \times 2$ | 6 | 4 | $2 \times 4$ |
| FLOP/cycle/core                 | 4 | 4 | 4 (SSE) | 4 |
| Peak Perf. [GFLOP/s]            | 102.4 | 128.16 | 56.0 | 147.2 |
| L1 [KiB]                        | 32 | 32 | 32 | 64 |
| L2 [KiB]                        | 6144 | 256 | 256 | 512 |
| L3 [MiB]                        | – | 12 | 8 | 5 |
| Sockets                         | 2 | 2 | 1 | 2 |
| NUMA domains                    | 1 | 2 | 1 | 4 |
| STREAM Copy BW                  |   |   |   |   |
| Core [GB/s]                     | 5.4 | 10.1 | 17.5 | 12.3 |
| Socket [GB/s]                   | 7.2 | 21.0 | 18.4 | 25.0 |
| Node [GB/s]                     | 7.0 | 40.6 | 18.4 | 49.0 |
| Machine Balance $B_{machine}$ [byte/FLOP] | 0.09 | 0.32 | 0.33 | 0.33 |

Table 7: Used test bed for performance evaluation. Copy stream bandwidth measured with McCalpin's STREAM [12] but using non-temporal stores. For computing the machine balance $B_{machine}$ the STREAM copy bandwidth of the full node was used.

The "Harpertown" system is a UMA system with two quad-core Intel Xeon X5482 CPUs at 3.2 GHz. Each core has a separate 32 KiB L1 cache and two cores each share a 6 MiB L2 cache.

The "Westmere" system is a two socket ccNUMA systems with hexa-core Intel Xeon X5650 CPUs supporting SMT and operating at 2.67 GHz. Each core has 32 KiB L1 data cache and 256 KiB L2 cache. Each chip has a 12 MiB L3 cache shared by the six cores.

The "SandyBridge" system is a Desktop-like node based on a quad-core Intel Xeon E3-1280 CPU at 3.5 GHz having only one NUMA domain, which makes the system behaving like a UMA system. All cores have a separate 64 KiB L1 and a separate 512 KiB L2 cache. The 8 MiB L3 cache is shared by all cores. With AVX instructions, the peak performance of SandyBridge is twice the one reported in Tab. 7 with SSE. However, the AVX extension of the instruction set is not used in the present study.

The two-socket "MagnyCours" system uses eight-core AMD Opteron 6134 CPUs running at 2.30 GHz. Each CPU package comprises two dies with four cores each. The L1 data cache with 64 KiB and the L2 cache with 512 KiB are separate for each core and the 5 MiB L3 cache is shared by all four cores on a die. Each die forms its own NUMA domain resulting in four domains for the whole system.

| propagation | full array | | | indirect addressing | | | |
| model | PDFs | $\frac{\text{bytes}}{\text{LUP}}$ | $B_{code}$ | PDFs | IDXs | $\frac{\text{bytes}}{\text{LUP}}$ | $B_{code}$ |
| --- | --- | --- | --- | --- | --- | --- | --- |
| TS    | 114 | 912 | 4.56 | 114 | 36 | 1056 | 5.28 |
| OS    | 57  | 456 | 2.28 | 57  | 18 | 528  | 2.64 |
| OS-NT | 38  | 304 | 1.52 | 38  | 18 | 376  | 1.88 |
| CG    | 56  | 448 | 2.24 | 56  | 18 | 520  | 2.60 |
| SWAP  | 56  | 448 | 2.24 | 114 | 36 | 484  | 2.42 |
| AAP   | 38  | 304 | 1.52 | 38  | 9  | 340  | 1.70 |
| ET    | 38  | 304 | 1.52 | 38  | 10 | 344  | 1.72 |

Table 8: Code balance $B_{code}$ (in bytes/FLOP) for different implementations of the propagation step with a D3Q19 stencil when a full array or indirect addressing is used. Note: size of a PDF and an IDX element is eight byte and four byte, respectively.

## 9. Performance Model

To estimate the performance that can be expected from the different implementations of the propagation step, a simple memory bandwidth based performance model [13] is used. This type of prediction works for loop kernels if i) the performance of the code is limited by the memory bandwidth and not by the computational performance and ii) the cache bandwidth is much larger than the memory bandwidth. For modern multicore chips, the model usually works well only if several cores that share a memory bus are utilized, even with memory-bound code [14].

A code is memory bound if its *code balance* $B_{code}$ is larger than the *machine balance* $B_{machine}$ of the system it should run on.

The code balance is the ratio of bytes required compared to the computations performed by the code in the kernel:

$$B_{code} = \frac{\text{transferred bytes}}{\text{floating-point operations}} \quad \frac{\text{byte}}{\text{FLOP}}. \quad (1)$$

In our case, this number is obtained by dividing the data requirements for each of the implementations by the number of FLOPs required for a lattice node update, which in our case of D3Q19 TRT is about 200 FLOPs. The results for double precision (8-byte) PDFs and 32-bit integers for the index array can be seen in Tab. 8.

The *machine balance* $B_{machine}$ on the other hand is defined as follows:

$$B_{machine} = \frac{\text{peak memory bandwidth}}{\text{peak floating-point performance}} \quad \frac{\text{bytes/s}}{\text{FLOP/s}}, \quad (2)$$

where we again assume double precision floating-point arithmetic. Table 7 shows $B_{machine}$ for all machines in the test bed. The actual machine balance may be slightly worse on the Intel Westmere and SandyBridge platform since clock frequency may be higher depending on the computational load due to Intel's "Turbo Mode" feature.

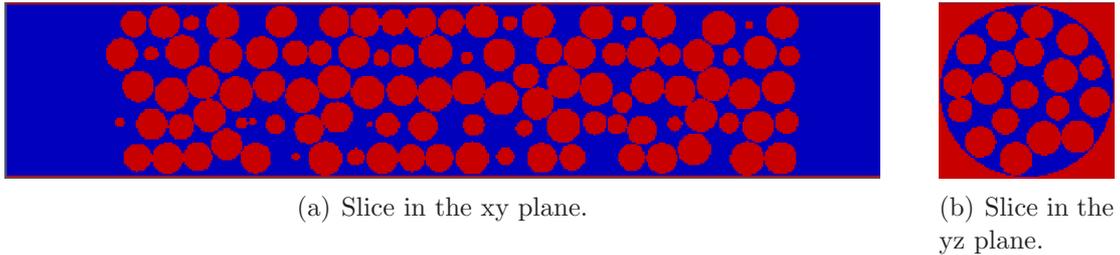

(a) Slice in the xy plane.

(b) Slice in the yz plane.

Figure 7: Slices of a packed bed reactor of spheres used as benchmark geometry. Blue indicates fluid cells and red solid cells.

Comparing $B_{code}$ of the implementations with $B_{machine}$ of the machines in the test bed shows that $B_{code} > B_{machine}$ always holds in our case. Thus, the memory bandwidth model can be applied to approximate an upper performance limit. Expected performance results are drawn as short lines in the performance graphs in Section 10. All implementations will still be memory bound even if SIMD vectorization is neglected and the system's lower scalar peak floating-point performance (not shown in Tab. 7) is used to calculate $B_{machine}$.

The code and machine balance can also be defined with single precision floating-point arithmetic. For the code balance this would reduce the required bytes by half, and the number of floating-point operations per core and cycle for the machine balance would be doubled. So the ratio between code and machine balance would stay the same.

All architectures in the test bed have separate floating-point units for addition and multiplication. Thus, the peak floating-point performance only can be achieved if two conditions are fulfilled: First, the code must have a balanced number of additions and multiplications to keep the units continuously busy. Second, for all floating-point operations packed arithmetic instructions must be used, i.e., packed SSE/AVX instructions on Intel and AMD machines.

## 10. Performance Results

We used the ILBDC (International Lattice Boltzmann Developer Consortium) flow solver [3], which is optimized for simulating the flow through porous media. It uses indirect addressing, a D3Q19 lattice, a two-relaxation-time (TRT) collision model [15], and double precision floating-point arithmetic. During the MPI-parallel simulation, each process holds the same number of fluid nodes [3, 16]. For compilation the Intel Fortran/C Compiler 11.1.075 was used together with Intel MPI 4.0.1.007. A packed bed of spheres (Fig. 7) and an empty channel were used as as benchmark geometries, both with the dimension of $500 \times 100 \times 100$ cells resulting in $\approx 2,100,000$ and $\approx 5,000,000$ fluid cells, respectively.

For performance evaluation, several setup strategies of the indirect addressing were benchmarked and the best was chosen for each machine and propagation model. The sparse representation setup for indirect addressing is explained in more detail in [3, 16].

Due to the structure of the existing code not all propagation models discussed in this paper have been implemented so far. Hence, only performance measurements of the one-step algorithm (OS) with push and pull, the one-step algorithm with pull and non-temporal stores (OS-NT, Pull) and the AA pattern (AAP) are discussed in the following. It is expected that esoteric twist performs similar to the AA pattern since its memory access behavior and the number of required bytes per LUP is the same. All shown results were obtained with the SoA data layout.

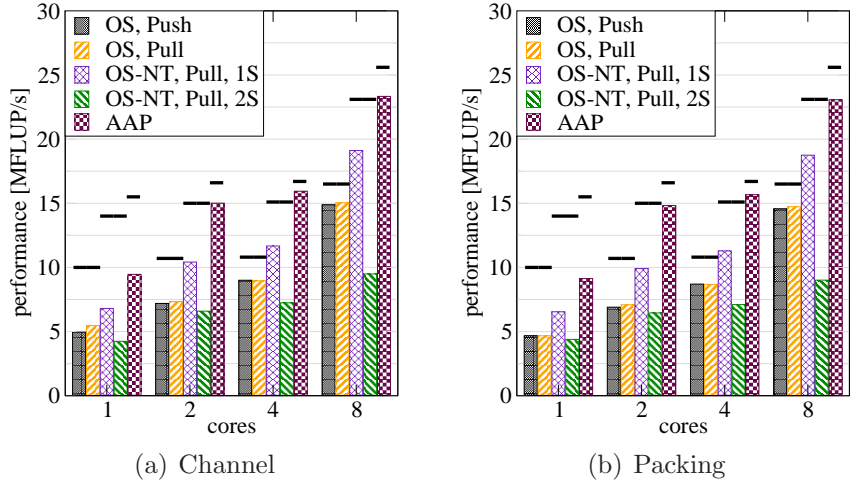

(a) Channel     (b) Packing

Figure 8: Performance of different propagation step implementations on **Harpertown** for the channel and packing benchmark geometry. Black lines indicate the maximum achievable performance expected by the bandwidth-based performance model in main memory.

*Harpertown.* Figure 8 shows the performance with the implemented propagation steps for the channel and packing geometry. The black lines mark the expected performance from the memory bandwidth model. If a socket is not fully utilized (one and two cores) only a fraction of the expected performance is reached. Using full sockets instead (four and eight cores), however, the performance model is quite accurate. Thus, the LBM implementation cannot saturate the memory bandwidth with a single core in contrast to the much simpler STREAM benchmark.

OS with push and pull shows no significant difference in performance. Interestingly OS-NT with $q$ store loops (OS-NT, Pull, 1S) performs significantly better than the version with $q/2$ store loops (OS-NT, Pull, 2S). In the latter version the compiler generates only one non-temporal store. The mix of temporal and non-temporal stores in one loop seems counterproductive on this microarchitecture.

*Westmere.* As expected from a ccNUMA system, performance scales across sockets (see Fig. 9). The OS algorithm achieves around 80 % of the expected performance on a full socket, independent of whether the push or the pull scheme is used. OS-NT achieves better performance than OS, but sustains less than 80 % of the expected value. The reason is that NT stores make less effective use of the memory interface on Intel architectures; the gain from the saved write-allocate transfer outweighs this drawback, however. Best performance is again achieved with AAP. A sustained performance of more than 100 MLUPs for D3Q19/TRT in double precision on a single two-socket compute node is remarkable.

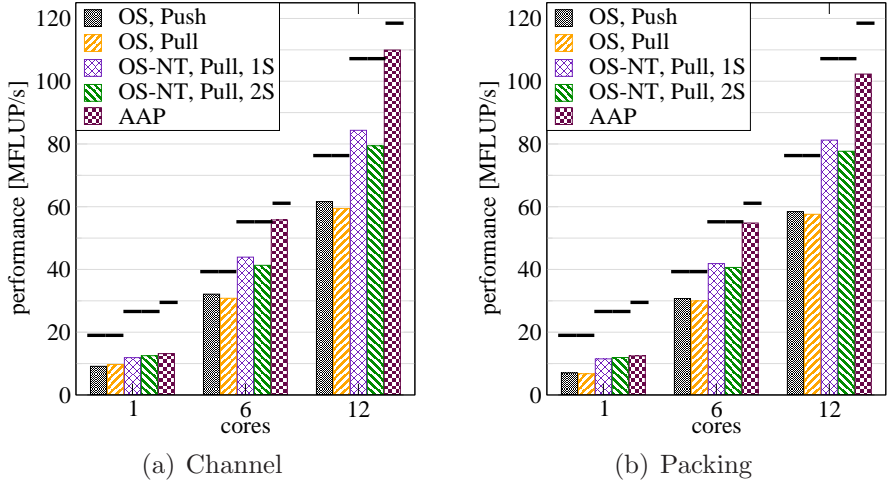

(a) Channel
(b) Packing

Figure 9: Performance of different propagation step implementations on **Westmere** for the channel and packing benchmark geometry. Black lines indicate the maximum achievable performance expected by the bandwidth-based performance model in main memory.

*SandyBridge.* For SandyBridge we expect about 45% of the performance of a Westmere node. None of the implemented algorithms fully meets this expectation. It is remarkable that there is no significant gain of AAP over OS-NT Pull 1S. As of now we do not have a conclusive explanation for those peculiarities.

*MagnyCours.* The performance scales (Fig. 11) across all four NUMA domains. Only around 65 % of the expected performance is reached by all implementations and APP is only at the level of OS-NT Pull 2S. This suggests that for this architecture the simple performance model does not work. One explanation might be the exclusive cache architecture [14] of the MagnyCours CPU, but confirmation requires a more detailed investigation.

## 11. Summary

The performance of a lattice Boltzmann based flow simulation is heavily influenced by the chosen implementation of the propagation step. Implementations like the one-step algorithm are simple to implement but easily outperformed by more advanced techniques like usage of non-temporal stores or the AA pattern. Special care has to be taken of machine-dependent properties like correct padding.

In this paper we only studied the single-node performance. However, a good single-node baseline is essential for any massively parallel simulation.

*Acknowledgements.* We are thankful for lively discussions with Jan Treibig and the careful proofreading by Rasa Mabande. The hospitality of the Intel Exascale Laboratory at University of Versailles Saint-Quentin-en-Yvelines while preparing parts of this paper is gratefully acknowledged by the first author (MW). This work was supported by BMBF under grant No. 01IH08003A (project SKALB).

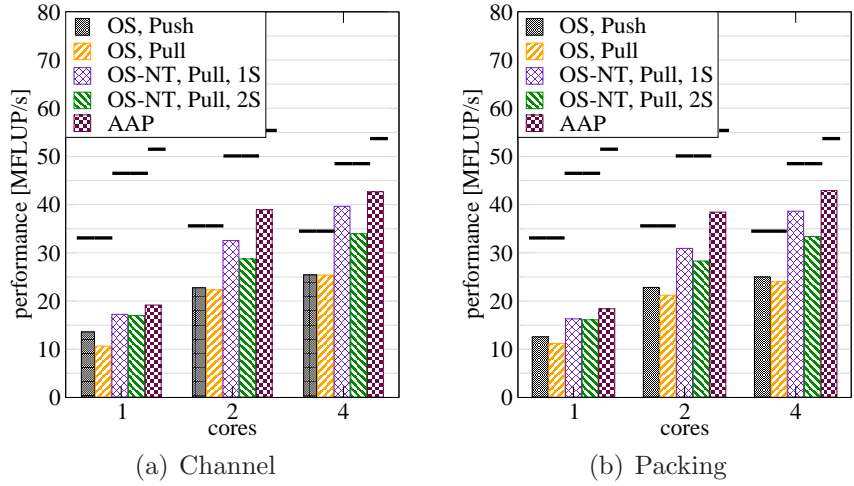

Figure 10: Performance of different propagation step implementations on **SandyBridge** for the channel and packing benchmark geometry. Black lines indicate the maximum achievable performance expected by the bandwidth-based performance model in main memory.

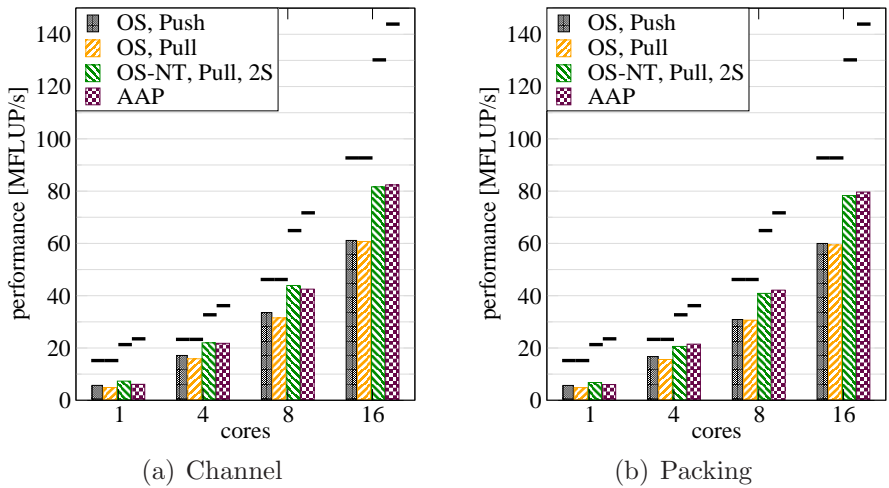

Figure 11: Performance of different propagation step implementations on **MagnyCours** for the channel and packing benchmark geometry. Black lines indicate the maximum achievable performance expected by the bandwidth-based performance model in main memory.

**References**

[1] S. Succi, The Lattice Boltzmann Equation for Fluid Dynamics and Beyond, Oxford University Press, 2001.
[2] C. Pan, J. F. Prins, C. T. Miller, A high-performance lattice Boltzmann implementation to model flow in porous media, Computer Physics Communications 158 (2004) 89–105.
[3] T. Zeiser, G. Hager, G. Wellein, Benchmark analysis and application results for lattice Boltzmann simulations on NEC SX vector and Intel Nehalem systems, Parallel Processing Letters 19 (2009) 491–511.
[4] T. Pohl, M. Kowarschik, J. Wilke, K. Iglberger, U. Rüde, Optimization and profiling of the cache performance of parallel lattice Boltzmann codes, Parallel Processing Letters 13 (2003) 549–560.
[5] K. Mattila, J. Hyväluoma, T. Rossi, M. Aspnäs, J. Westerholm, An efficient swap algorithm for the lattice Boltzmann method, Computer Physics Communications 176 (2007) 200–210.
[6] J. Latt, How to implement your DdQq dynamics with only q variables per node (instead of 2q), Technical Report, Tufts University, Medford, USA, 2007.
[7] P. Bailey, J. Myre, S. Walsh, D. Lilja, M. Saar, Accelerating lattice Boltzmann fluid flow simulations using graphics processors, in: International Conference on Parallel Processing 2009 (ICPP'09).
[8] M. Schönherr, M. Geier, M. Krafczyk, 3D GPGPU LBM implementation on non-uniform grids, in: International Conference on Parallel Computational Fluid Dynamics 2011 (Parallel CFD 2001).
[9] J. Linxweiler, Ein integrierter Softwareansatz zur interaktiven Exploration und Steuerung von Strömungssimulationen auf Many-Core-Architekturen, Ph.D. thesis, Fakultät Architektur, Bauingenieurwesen und Umweltwissenschaften, TU-Braunschweig, 2011.
[10] G. Wellein, T. Zeiser, G. Hager, S. Donath, On the single processor performance of simple lattice Boltzmann kernels, Comput. Fluids 35 (2006) 910–919.
[11] G. Wellein, G. Hager, T. Zeiser, M. Wittmann, H. Fehske, Efficient temporal blocking for stencil computations by multicore-aware wavefront parallelization, in: Proceedings of the 33rd IEEE International Computer Software and Applications Conference (COMPSAC 2009), pp. 579–586. doi.ieeecomputersociety.org/10.1109/COMPSAC.2009.82.
[12] J. D. McCalpin, Memory bandwidth and machine balance in current high performance computers, IEEE Computer Society Technical Committee on Computer Architecture (TCCA) Newsletter (1995) 19–25. http://www.cs.virginia.edu/stream/.
[13] G. Hager, G. Wellein, Introduction to High Performance Computing for Scientists and Engineers, CRC Press, 2010. ISBN 978-1439811924.
[14] J. Treibig, G. Hager, Introducing a performance model for bandwidth-limited loop kernels, in: R. Wyrzykowski, J. Dongarra, K. Karczewski, J. Wasniewski (Eds.), Parallel Processing and Applied Mathematics, volume 6067 of *Lecture Notes in Computer Science*, Springer-Verlag, Berlin, Heidelberg, 2010, pp. 615–624. dx.doi.org/10.1007/978-3-642-14390-8_64.
[15] I. Ginzburg, J.-P. Carlier, C. Kao, Lattice Boltzmann approach to Richards' equation, in: C. T. Miller (Ed.), Computational Methods in Water Resources: Proceedings of the CMWR XV, June 13–17, 2004 Chapel Hill, NC, USA, Elsevier, Amsterdam, 2004, pp. 583–597.
[16] M. Wittmann, T. Zeiser, G. Hager, G. Wellein, Domain decomposition and locality optimization for large-scale lattice Boltzmann simulations, Submitted to ParCFD-2011 Special Issue of Computer & Fluids (2011).